\documentclass[3p,12pt]{elsarticle}
\usepackage{amssymb}
\usepackage{float,epsfig,latexsym,dcolumn,lscape,longtable,setspace}
\usepackage{makeidx}
\usepackage{graphicx}
\usepackage{amsmath}
\usepackage{natbib}
\usepackage{fancyhdr}
\usepackage{subfigure}
\usepackage{threeparttable}
\usepackage{booktabs}
\usepackage{microtype}
\usepackage{amsmath}

\makeatletter
\def\ps@pprintTitle{  \let\@oddhead\@empty
  \let\@evenhead\@empty
  \let\@oddfoot\@empty
  \let\@evenfoot\@oddfoot
}
\makeatother

\newcolumntype{d}[1]{D{.}{.}{#1}}
\newcolumntype{p}[1]{D{(}{(}{-1}}

\bibpunct{(}{)}{;}{a}{,}{,}

\begin{document}

\begin{frontmatter}

\title{RHOMOLO: A Dynamic Spatial General Equilibrium Model for
Assessing the Impact of Cohesion Policy\tnoteref{thanks}}

\tnotetext[thanks]{The authors acknowledge valuable contributions from
Janos Varga as well as from participants to various seminars at the European
Commission and ERSA conferences. The views expressed are
the authors' and do not necessarily correspond to those of the European
Commission.}

\author[IPTS]{Andries Brandsma}
\author[IPTS]{d'Artis Kancs\corref{dartis}}
\ead{d'artis.kancs@ec.europa.eu}
\cortext[dartis]{Corresponding author. European Commission, DG JRC, IPTS, E-41092 Seville, Spain.}
\author[REGIO]{Philippe Monfort}
\author[REGIO]{Alexandra Rillaers}

\address[IPTS]{European Commission, DG Joint Research Centre, Seville, Spain}
\address[REGIO]{European Commission, DG Regional and Urban Policy, Brussels, Belgium}

\begin{abstract}
The paper presents the newly developed dynamic spatial general equilibrium model of European Commission, RHOMOLO.  The model incorporates several elements from economic geography in a novel and theoretically consistent way. It describes the location choice of different types of agents and captures the interplay between agglomeration and dispersion forces in determining the spatial equilibrium. The model is also dynamic as it allows for the accumulation of factors of production, human capital and technology. This makes RHOMOLO well suited for simulating policy scenario related to the EU cohesion policy and for the analysis of its impact on the regions and the Member States of the union.
\end{abstract}

\begin{keyword}
Economic modelling, spatial dynamics, policy impact
assessment, regional development, economic geography, spatial equilibrium, DSGE. \\
\textit{JEL\ code: C63, C68, D58, F12, H41, O31, O40, R13, R30, R40.}
\end{keyword}

\end{frontmatter}

\vfill

\begin{center}
\today
\end{center}

\renewcommand{\thefootnote}{\arabic{footnote}} \setcounter{footnote}{0} %
\thispagestyle{empty}\addtocounter{page}{-1} \onehalfspacing
\newpage

\section{Introduction}

\subsection{Why developing a new model?}

For years, the Directorate General for Regional and Urban Policy of the
European Commission had used economic models for analysing the impact of
cohesion policy programmes. In particular, DG REGIO extensively relied on
two models for the simulation of scenarios related to cohesion policy:
HERMIN and QUEST. HERMIN was initially developed by scholars in the 1980's
and has been regularly upgraded since then. QUEST is the model developed and
used by the Directorate General for Economic and Financial Affairs %
\citep{VargaintVeld11}. It adopts the most recent practices in DSGE
modelling, which is notably reflected in its high level of micro-foundations.

However, given that both these models produce results at the national level,
it was felt that DG REGIO should extend its analytical capacities to also
cover the regional level. After an in-depth literature review, it appeared
that none of the existing models could fully respond to the need of DG REGIO
which hence decided to develop its own regional model.

The objective was to build a dynamic spatial general equilibrium model which
would be suited for analysing the impact of cohesion policy at the NUTS 2
level, i.e. the most relevant geographical level for the policy.\footnote{%
In some cases, NUTS 2 regions are relatively small (like for instance some
German L\"{a}nders) and the NUTS 1 level was then considered as more
appropriate.} In order to cover the needs of DG REGIO, the model had to
include several features. In particular, since cohesion policy mostly
supports investments aiming at fostering economic growth in EU regions, the
model should be well suited to capture the impact of the policy on the main
engines of endogenous growth. At the same time, it should account for local
specificities which may affect the dynamics of the regional economies
(factor endowment, accessibility, etc.).

Finally, the model should incorporate regional linkages in the line of New
Economic Geography and be capable of simulating the impact of policy shocks
on the spatial equilibrium. Accordingly, the model should incorporate
various agglomeration and dispersion forces as well as other possible
sources of spatial spill-over and interdependencies.

First, a prototype of the model was elaborated by a private consultant (TNO)
contracted by DG REGIO.\footnote{%
See \cite{FerraraIvanovaKancs10} for a formal description of the prototype
model.} The prototype was then passed on to DG REGIO and Directorate General
Join Research Centre (DG JRC) which developed a dynamic spatial general
equilibrium model covering the EU-27 at NUTS 2 level. The model has been
named RHOMOLO, standing for Regional HOlistic MOdeL.

\subsection{Main features of RHOMOLO}

The domestic economy (which corresponds to the EU) consists of $R_{-1}$
regions $r=1,\ldots ,R_{-1}$, which are included into $M$ countries $%
m=1,\ldots ,M$. Each region is inhabited by $H_{r}$ households which are
immobile. They partly determine the size of the regional market.\footnote{%
Labour mobility can be introduced through a module which extends this core
version of the model with a more sophisticated specification of the labour
market. This is described in \cite{BrandsmaKancsPersyn14}.} The income of
households consists in labour revenue (wages), capital revenue and
government transfers. It is used to consume final goods, pay taxes and
accumulate savings.

The final goods sector includes $s=1,\ldots ,S$ different economic sectors
in which firms operate under monopolistic competition \`{a} la %
\citet{DixitStiglitz77}. Each firm produces a differentiated variety which
is considered as an imperfect substitute to the other by households and
firms. The number of firms in sector $s$ and region $r$ is denoted by $%
N_{s,r}$. It is large enough so that strategic interactions between firms is
negligible. Goods are either consumed by households or used by other firms
as intermediate inputs or as investment goods. The number of firms in each
region is endogenous and to a large extent determines the spatial
distribution of economic activity.

The rest of the world is introduced in the model as a particular region
(indexed by $R $) and particular sector (indexed by $S$). Sector $S$ differs
from domestic sectors in that it only has one variety which is exclusively
produced in region $R$. Formally, we have $N_{S,r}=0$ and $N_{s,R}=0$ for
all $r$ and $s$; and $N_{S,R}=1$. The foreign variety of final goods is used
as the numeraire.

Trade between (and within)\ regions is costly, implying that the shipping of
goods between (and within)\ regions entails transport costs which are
assumed to be of the iceberg type, with $\tau _{s,r,q}>1$ representing the
quantity of sector's $s$ goods which needs to be sent from region $r$ in
order to have one unit arriving in region $q$ (see for instance %
\citet{krugman1991increasing}). Transport costs are assumed to be identical
across varieties but specific to sectors and trading partners. They are
related to the distance separating regions $r$ and $q$ but can also depend
on other factors, such as transport infrastructure or national borders.
Finally, transport costs can be asymmetric (i.e. $\tau _{s,r,q}$ may differ
from $\tau _{s,q,r}$). They are also assumed to be positive within a given
region (i.e. $\tau _{s,r,r}\neq 1$) which captures, among others, the
distance between customers and firms within the region.

The description of technological change directly follows \citet{Romer1990}
and \citet{jones95}. In their production process, final goods firms use
durable goods. Each variety of durable goods is produced by a specific
(durable goods) firm and traded on a regional market whose structure is
monopolistic competition. In order to start operating, each durable goods
firm must acquire a design from a research sector which uses human capital
and existing knowledge to produce new designs.

The structure of the labour market is also monopolistic competition. Each
household supplies a specific variety of low, medium and high skilled labour
services to firms which are considered as imperfect substitutes to the ones
offered by other households. Changing wages is assumed to be costly which
introduces nominal wage rigidity in the model.\footnote{%
Assuming monopolistic competition both on the goods and the labour markets
is in line with models such as \citet{Blanchard-Kiyotaki1987}. Such
specification allows to focus both on price and wage decisions and in
particular to introduce nominal rigidities in the model. In addition, as
underlined by \citet{Manning2010}, monopolistic competition is a very simple
manner to capture the idea of thick labour markets, namely that the high
density of workers raises productivity, thereby making large labour markets
more attractive for firms. This enriches the the description of the economic
geography in the mode.}

Finally, in each country there is a public sector which levies taxes on
consumption and on the income of local households. It provides public goods
in the form of public capital which is necessary for the operation of firms.
It also subsidises the private sector, including the production of R\&D and
innovation, and influences the capacity of the educational system to produce
human capital.

The detailed regional and sectoral dimensions of RHOMOLO implies that the
number of (non-linear) equations to be solved simultaneously is relatively
high. Therefore, in order to keep the model manageable from a computation
point of view, its dynamics is kept relatively simple. Three types of
factors (physical capital, human capital and knowledge capital) as well as
three types of assets (equities, domestic government bonds and foreign
bonds) are accumulated between periods. Agents are assumed to save a
constant fraction of their income each period and to form their expectations
based only on the current and past states of the economy. The dynamics of
the model is then described as in a standard Solow model, i.e. a sequence of
static sub-models that are linked between periods by the laws of motion
determining the time path of some key stock variables.

The model includes several agglomeration and dispersion forces determining
the location choice of firms. Those include backward (firms prefer to have
good access to output markets) and forward linkages (firms prefer to have
good access to input markets) as well as Marshallian technological
spill-over. Dispersion forces relate to competition on the goods market as
well as competition for local labour.

This paper aims at presenting the theoretical specifications underlying
RHOMOLO in order to document and clarify the main assumptions and
micro-founded mechanisms it contains. Section 2 details the behaviour of
households while section 3 focuses on firms in the final goods sector. It
also describes how the interplay between the R\&D and the durable goods
sector leads to technological progress. Section 4 is devoted to the public
sector and explains how policy interventions are introduced in the model.
Section 5 lays down the conditions to clear the product, labour and
financial markets and elaborates on the notion of spatial equilibrium in the
model. Finally, section 7 concludes.

\section{Households}

Households make decisions about consumption, savings and labour supply. Each household supplies a differentiated
variety of labour which contains a low, medium and high skilled component.
Let $e=lo,me,hi$ denote the low-, medium- and high-skilled component
respectively. Preference of households is represented by a utility function
which is additively separable in consumption and leisure:
\begin{equation*}
U \left(C_{h,q} ; \sum_{e=lo,me,hi}V(1-l_{h,e,q})\right) = C_{h,q} +
\sum_{e=lo,me,hi}V(1-l_{h,e,q})
\end{equation*}%
where $C_{h,q}$ is the consumption of final goods by household $h$ in region
$q$ is and $l_{h,e,q}$ is the labour of type $e$ it supplies. We assume that
the sub-utility with respect to leisure takes the form of a CES with a
standard labour supply elasticity ($\kappa$) and a skill specific weight ($%
\omega _{e}$) on leisure in order to capture differences in participation to
the labour market between skill groups. We have
\begin{equation*}
\sum_{e}V(1-l_{h,e,q})=\sum_{e}\frac{\omega _{e}}{1-\kappa } \,
(1-l_{h,e,q})^{1-\kappa}
\end{equation*}

The budget constraint of household $h$, $q$ can be written as
\begin{equation}
P_{q}^{c} C_{h,q} \leq (1-s) \, YC_{h,q}  \label{BC}
\end{equation}

where $P_{q}^{c}$ is the price of final goods in region $q$, $YC_{h,q}$ is
disposable income and $s$ is the constant saving rate which is common to all
households.

Disposable income is the sum of labour and capital income net of wage
adjustment cost plus government transfers net of taxes:
\begin{equation*}
YC_{h,q}=\sum_{e} (1-t_{m}^{w}) \, w_{h,e,q} \, l_{h,e,q} - \Gamma_{w}
(w_{h,e,q}) + (1-t_{m}^{\pi }) \, KI_{h,q}+\frac{TR_{H,m}}{%
\sum_{r=1}^{R_{m}}H_{r}}
\end{equation*}
where $w_{h,e,q}$ is the wage paid to household $h$, $q$ for its skill level
$e$, $KI_{h,q}$ is capital income, and $TR_{H,m}$ denotes government
transfers to households in country $m$.$\Gamma_{w} (w_{h,e,q})$ denotes the
wage adjustment cost (see below).

Capital income corresponds to the returns linked to the holding of three
different types of assets: equities (i.e. liability against the durable
goods firms in the $R_{-1}$ regions of the domestic economy), domestic bonds
(i.e. liability against the $M$ governments of the domestic economy) and
foreign bonds (i.e. liability against the rest of the world). Let $%
B_{h,q}^{k,v,r}$, $B_{h,q}^{G,m}$, $B_{h,q}^F$ denote the stock of these
assets held by the household respectively in firm $v$ of region $r$, in
government bonds of country $m$ and in foreign bonds. The associated returns
are respectively $r_{v,r}^{k}$, $r_{m}^{G}$, $r^F$. The holding of equities
also gives right to a share of the durable goods firms profit. Finally, we
assume that the firms in the final goods sector are also owned by households
who share their profits. Capital income then reads
\begin{equation}
KI_{h,q} = \sum_{r=1}^{R-1} \sum_{v=1}^{A_r} r_{v,r}^{k} B_{h,q}^{k,v,r} +
\sum_{m=1}^{M} r_{m}^{G} B_{h,q}^{G,m} + r^F B_{h,q}^F + \sum_{r=1}^{R-1}
\sum_{v=1}^{A_r} s_{h,q}^{k, v, r} \pi_{v,r} + \frac{1}{H} \pi^{FG}
\label{KI}
\end{equation}%
where ${A_r}$ is the number of durable goods firms in region $r$, $\pi_{v,r}$
is the profit of the durable goods firm $v$ in region $r$, $%
H=\sum_{r=1}^{R-1} H_r$ is the domestic population and $\pi^{FG}$ is the sum
of profits of the final goods sector. $s_{h,q}^{k, v, r} = B_{h,q}^{k,v,r} /
a_{v,r}$ is the share of total assets issued by firm $v, r$ ($a_{v,r}$) held
by the household. Except for $H$, these variables are endogenous and their
determination is described in the next sections of the paper.

Finally, wages are subject to convex adjustment cost:
\begin{equation*}
\Gamma_{w} (w_{h,e,q}) = \sum_{e} \frac{\gamma_{w}}{2} \, l_{h,e,q} \, \frac{%
\Delta w_{h,e,q}^2}{w_{h,e,q}}
\end{equation*}

The optimisation problem of the household is solved by maximising the
associated Lagrangian
\begin{equation}
L = C_{h,q} + \sum_{e} \frac{\omega _{e}}{1-\kappa }(1-l_{h,e,q})^{1-\kappa
} - \lambda \left( P_{q}^{c} \, C_{h,q} - (1-s) \,YC_{h,q}\right)
\label{Lagrangian}
\end{equation}
with respect to consumption, $C_{h,q}$, and labour supply, $l_{h,e,q}$.

\subsection{Consumption}

First order conditions to that optimisation problem imply that the aggregate
consumption level is directly related to disposable income:
\begin{equation*}
C_{h,q} = \frac{(1-s) \, YC_{h,q}}{ P_{q}^{c}}
\end{equation*}

Households consume all varieties of final goods available in the economy. In
order to represent love for varieties, $C_{h,q}$ is assumed to take the form
of a CES sub-utility function defined as\footnote{%
Distribution parameters are introduced in the CES in order to calibrate the
model (see for instance \citet{Anderson-vanWincoop2003} or %
\citet{Balisteri2011}). These parameters are not included here for the sake
of simplicity.}
\begin{equation}
C_{h,q}=\left( \sum_{r=1}^{R}\sum_{s=1}^{S} \sum_{i=1}^{N_{s,r}}\left(
c_{h,q}^{i,s,r}\right) ^{\theta }\right) ^{\frac{1}{\theta }}  \label{cons1}
\end{equation}
where $c_{h,q}^{i,s,r}$ is the consumption of variety $i$ of sector $s$
produced in regions $r$ and $\beta _{s}$ is the weight given to sector $s$
in the household's preference.\footnote{%
The model as coded incorporates a nested CES utility function to allow for
different elasticities of substitution between varieties of a given sector
on the one hand and sectors on the other hand. This feature is not
introduced here to simplify notations.}

Household $h, q$ chooses a consumption bundle in order to maximise (\ref%
{cons1}) subject to the following constraint:
\begin{equation*}
\sum_{r=1}^{R}\sum_{s=1}^{S}\sum_{i=1}^{N_{s,r}}\tau _{s,r,q}\left(
1+t_{s,m}^{c}\right) p_{i,s,r} \, c_{h,q}^{i,s,r}=(1-s)\,YC_{{h,q}}
\end{equation*}%
where $p_{i,s,r}$ is the price of variety $i$,$s$,$r$, $\tau _{s,r,q}$ is
trade cost from region $r$ to region $q$, and $t_{s,m}^{c}$ is the tax rate
applied to consumption of sector $s$ goods in country $m$ (where region $q$
is assumed to be located).

The price of variety $S, R$ produced in the rest of the world is assumed to
be exogenous to the domestic economy, i.e. $p_{S,R}=\bar{p}_{S,R}$. We also
assume that foreign households have the same type of preference regarding
domestic goods and that the share of their disposable income devoted to the
consumption of domestic goods is fixed.

Solving this problem leads to the following demand for variety $i$,$s$,$r$:
\begin{equation}
c_{h,q}^{i,s,r}=\left( \frac{\tau _{s,r,q}\left( 1+t_{s,m}^{c}\right)
p_{i,s,r}}{\beta _{s}P_{q}^{c}}\right) ^{\frac{1}{\theta -1}}\frac{%
(1-s)\,YC_{h,q}}{P_{q}^{c}}  \label{eq.consumption2}
\end{equation}%
where $P_{q}^{c}$ is the following CES price index:
\begin{equation}
P_{q}^{c}=\left( \sum_{r=1}^{R}\sum_{s=1}^{S}\sum_{i=1}^{N_{s,r}}\beta _{s}^{%
\frac{1}{1-\theta }}(\tau _{s,r,q}\left( 1+t_{s,m}^{c}\right) p_{i,s,r})^{%
\frac{\theta }{\theta -1}}\right) ^{\frac{\theta -1}{\theta }}
\label{eq.consumer.price.index}
\end{equation}

According to (\ref{eq.consumption2}), demand for variety $i, s, r$ is a
fraction of real income spent on final goods. This fraction decreases with
the relative price of this variety, the relevant transport cost and tax
rate, while it increases with relative preference for sector $s$.

\subsection{Labour supply}

Each household decides which fraction of its time endowment will be devoted
to leisure on the one hand and to labour in each types of skill on the other
hand. Labour markets are characterised by monopolistic competition where,
within each skill group, labour supplied by a particular household
corresponds to a variety which is an imperfect substitute to the others.
Maximising (\ref{Lagrangian}) with respect to $l_{h,e,q}$, we obtain the
following wage setting rule:

\begin{equation}
\omega _{e} \, (1-l_{h,e,q})^{-\kappa } \frac{1}{\eta} = (1-t_{m}^w) \frac{%
w_{h,e,q}}{P_{q}^{c}}  \label{labour-supply}
\end{equation}
with
\begin{equation*}
\eta = \left[\sigma \, (1-s) - \frac{\gamma_w \, (\sigma-1) \,
\pi_{h,e,q}^{w}}{(1-t_{m}^w)} \right]
\end{equation*}

The real wage is set as a mark-up, $\frac{1}{\eta}$, over the reservation
wage (i.e. the marginal utility of leisure divided by the marginal utility
of consumption). The mark-up depends on the elasticity of substitution
between the different varieties of labour in the firms production function, $%
\sigma$ (see below). Because adjusting wages is costly, the mark-up also
depends on the level of wage inflation, $\pi_{h,e,q}^{w} = \Delta
w_{h,e,q}/w_{h,e,q}$, which implies that wages only adjust slowly to
variation in prices.

It is further assumed that the household accumulates skill specific human
capital, $b_{h,e,q}$, according to

\begin{equation}
\Delta b_{h,e,q}= b_{h,e,q}(e^\Lambda_{h,e,q}-1) - \delta_{HC} \, b_{h,e,q},
\end{equation}

allowing to offer $b_{h,e,q}$ efficiency units. $\Lambda$ represents the
amount of time a household spends on education while $\delta_{HC}$ is the
depreciation rate for human capital.

\section{Firms}

\subsection{Final goods firms}

Final goods firms produce horizontally differentiated varieties of final
goods. The final goods sector is
characterised by monopolistic competition. It includes $S$ sectors in which
each firm produces a variety which is an imperfect substitute to the others.
The production function firms is of the Leontieff type. The arguments are
the quantities of intermediate inputs bought from all sectors and a
Cobb-Douglas aggregate of the factors used in the production process, i.e.
labour and durable goods:
\begin{equation}
X_{i,s,r}=\min \{y_{i,s,r},a_{s}^{1}X_{i,s,r}^{1},\ldots
,a_{s}^{u}X_{i,s,r}^{u},\ldots ,a_{s}^{S}X_{i,s,r}^{S}\}
\label{eq.prod.function}
\end{equation}%
where $X_{i,s,r}$ is the quantity produced by the firm producing variety $i$
of sector $s$ located in region $r$ , $y_{i,s,r}$ is its value added, $%
X_{i,s,r}^{u}$ is an index of the intermediate inputs from sector $u$ and $%
a_{s}^{u}$ the associated technical coefficient, assumed to be common to all
firms in sector $s$ in country $m$, independently of their location within
the country. The index $X_{i,s,r}^{u}$ is a CES aggregate of varieties
produced in sector $u$:
\begin{equation*}
X_{i,s,r}^{u}=\left( \sum_{q=1}^{R}\sum_{j=1}^{N_{u,q}}\left( {%
x_{i,s,r}^{j,u,q}}\right) ^{\theta }\right) ^{\frac{1}{\theta }}
\end{equation*}%
with $\theta \in (0,1)$.

The firm's value added is a Cobb-Douglas of the two factors used in the
production process:
\begin{equation}
y_{i,s,r}=Z_{i,s,r}^{\alpha _{s}}\,L_{i,s,r}^{1-{\alpha _{s}}%
}\,KG_{r}^{\alpha _{G}}-FC_{i,s,r}  \label{eq.value.added}
\end{equation}%
where $Z_{i,s,r}$ and $L_{i,s,r}$ are CES aggregates of the varieties of
durable goods and of the various types of labour --low-, medium- and
high-skilled-- used by the firm.\footnote{%
The firm uses effective units of labour which includes both physical units
of labour and the associated human capital.} Let $KG_{r}$ denote the stock
of public capital available in region $r$ which is assumed to affect
positively total factor productivity.\footnote{%
Note that according to this specification, each firm can benefit from the
whole stock of public capital available in the region where it is located.
This reflects the public good nature of public capital and in particular
that it is non-rivalrous. We also assume it is non-excludable in that its
use by firms does not incur direct payment but only indirect ones (the
provision of pubic capital is financed by taxes) which are not internalised
by the firm.} The firm also supports a fixed cost, $FC_{i,s,r}$, made of
some of the firm's value added. \footnote{%
The model also incorporates a fixed cost in terms of labour. It is not
included here in order to simplify the presentation.} Finally, it benefits
from subsidies of the national government ($Sub_{m}^{i,s,r}$) and of the EU (%
$Sub_{EU}^{i,s,r}$).

Durable goods and labour are assumed to be spatially immobile which implies
that firms in regions $r$ can only obtain those two factors on the local
market. Moreover, we assume that durable goods are not subject to internal
transport costs (i.e. $\tau _{Z,r,r}=1$). The respective CES indices read
\begin{eqnarray*}
Z_{i,s,r} &=&\left( \sum_{v=1}^{A_{r}}({z_{i,s,r}^{v,r}})^{\rho }\right) ^{%
\frac{1}{\rho }} \\
L_{i,s,r} &=&\left( \sum_{e=lo,me,hi} \gamma _{e} \sum_{h=1}^{H_{r}}{%
(b_{h,e,r} \, l_{i,s,r}^{h,e,r}})^{\sigma }\right) ^{\frac{1}{\sigma }}
\end{eqnarray*}%
where $\rho ,\sigma \in (0,1)$. Factor $\gamma _{e}$ accounts for difference
in labour productivity between low, medium and skilled labour, with $\gamma
_{lo}<\gamma _{me}<\gamma _{hi}$.

Profit maximisation leads the firm to set the output price as a mark-up over
marginal cost, where the mark-up depends on the elasticity of the total
demand it faces. This includes demand from households, from other firms for
intermediate inputs, from durable goods firms for investment goods and from
the government. Given our assumptions concerning the preferences of these
agents and the CES aggregates for intermediate inputs and for physical
capital (see below), the elasticity of total demand is $1/(\theta-1)$ and
the price-making rule is
\begin{equation}
p_{i,s,r} =\frac{MC_{i,s,r}}{\theta }  \label{PF}
\end{equation}

The marginal cost includes the cost of production factors and the cost of
intermediate inputs:
\begin{equation*}
MC_{i,s,r}=P_{i,s,r}^{y}+\sum_{u=1}^{S}{\ a_{s}^{u}\cdot P_{i,s,r}^{u}}
\end{equation*}%
where $P_{i,s,r}^{y}$ is the price of value added. Given the specification
adopted for valued added, $P_{i,s,r}^{y}$ is common to all firms in sector $%
s $ and region $r$ and corresponds to a Cobb-Douglas of the factors' price:
\begin{equation*}
P_{i,s,r}^{y}=KG_{r}^{-\alpha _{G}}\cdot \left( \frac{P_{i,s,r}^{z}}{\alpha
_{s}}\right) ^{\alpha _{s}}\cdot \left( \frac{W_{i,s,r}}{1-\alpha _{s}}%
\right) ^{1-{\alpha _{s}}}
\end{equation*}%
$P_{i,sr}^{u}$, $P_{i,sr}^{z}$ and $W_{i,s,r}$ are the price indices
corresponding to the CES aggregates respectively of intermediate inputs,
durable goods and labour varieties:
\begin{eqnarray}
P_{i,s,r}^{u} &=&\left( \sum_{q=1}^{R}\sum_{j=1}^{N_{u,q}}(\tau _{u,q,r}
\,p_{j,u,q})^{\frac{\theta }{\theta -1}}\right) ^{\frac{\theta -1}{\theta }}
\label{eq.intermed.price.index} \\
P_{i,s,r}^{z} &=&\left( \sum_{v=1}^{A_{r}} {p_{v,r}^{z}}^{\frac{\rho }{\rho
-1}}\right) ^{\frac{\rho -1}{\rho }}  \label{eq.specialis.price.index} \\
W_{i,s,r} &=&\left( \sum_{e={lo,me,hi}}\sum_{h=1}^{J_{r}}b_{h,e,r}^{\frac{%
\sigma}{1-\sigma }}w_{h,e,r}^{\frac{\sigma }{\sigma -1}}\right) ^{\frac{%
\sigma -1}{\sigma }}  \label{eq.labour.price.index}
\end{eqnarray}%
where $p_{j,u,q}$ is the price of variety $j, u, q$ of final goods, $%
p_{v,r}^{z}$ is the price of variety $v, r$ of durable goods and $w_{h,r,e}$
is the wage of household $h, r$ for his labour service of skill $e$.

We assume symmetry across firms (resp. households) in terms of the
technology (resp. preferences) which implies that the price (resp. wage) set
by each firm (resp. household) within one given region is the same.
Accordingly, one easily verifies that $P_{i,s,r}^{u}=P_{r}^{u}$ for all $i,
s $, $P_{i,s,r}^{z}=P_{r}^{z}$ for all $i, s$, $W_{i,s,r}=W_{r}$ for all $i,
s$, and $P_{i,s,r}^{y}=P_{s,r}^{y}$ for all $i$. Note that we also that
consumption taxes do not apply to intermediate inputs.

The demand of the firm for each variety of intermediate inputs, durable
goods and labour then take the following form, respectively:
\begin{eqnarray}
x_{i,s,r}^{j,u,q} &=&\left( \frac{ \tau_{u,q,r} \, p_{j,u,q}}{P_{r}^{u}}%
\right)^{\frac{1}{\theta -1}}X_{i,s,r}^{u}  \label{F} \\
z_{i,s,r}^{v,r} &=&\left( \frac{p_{v,r}}{P_{r}^{z}}\right) ^{\frac{1}{\rho -1%
}}Z_{i,s,r}  \label{Z} \\
l_{i,s,r}^{h,e,r} &=&\left( \frac{w_{h,e,r}}{b_{h,e,r}^{\sigma} \, W_{r}}%
\right) ^{\frac{1}{\sigma -1}}L_{i,s,r}  \label{l}
\end{eqnarray}

\subsection{National R\&D sectors}

There are $M$ national R\&D sectors which produce new designs $\Delta J_{m}$
using all varieties of skilled labour available on the national labour
market. The production process features
learning by doing, as labour productivity is positively related to the
pre-existing stock of designs. There are international technological
spill-overs in the sense that the national R\&D sector absorbs part of the
technology produced within the $M$ countries. Finally, the R\&D sector is
supported by the national government and the EU which provide subsidies, $%
Sub_{m}^{R\&D}$ and $Sub_{EU}^{R\&D}$, proportional to the production of new
designs. Following \cite{Romer1990}, the production function of the R\&D
sector of country $m$ reads
\begin{equation*}
\Delta J_{m}=(J^{\ast })^{\omega }\cdot J_{m}^{\zeta }\cdot L_{R\&D,m}^{hi}%
\hspace{0.25in}\omega ,\zeta <1
\end{equation*}%
where $J^{\ast }$ is the stock of design in the $M$ economies and $%
L_{R\&D,m} $ is a CES aggregate of the national skilled labour varieties
\begin{equation*}
L_{R\&D,m}=\left( \sum_{r=1}^{Rm}\sum_{h=1}^{H_{r}}{(b_{h,hi,r} \,
l_{R\&D}^{h,hi,r})}^{\sigma }\right) ^{\frac{1}{\sigma }}
\end{equation*}

Finally, designs are assumed to become obsolete after one period (i.e. the
depreciation rate on designs is set to 1) which implies that firms must
renew their licences every year to continue using the updated versions of
the designs. \footnote{%
In fact, this assumption is adopted in order to avoid introducing
inter-temporal decisions in the model and hence keep the description of its
dynamics simple.}

Perfect competition prevails on each national market for designs and firms
maximise profit by choosing the level of new designs and the corresponding
quantity of skilled labour employed in each variety:
\begin{equation*}
\Delta J_{m}=\left( \Omega \cdot \frac{P_{J,m} + Sub_{m}^{R\&D} +
+Sub_{EU}^{R\&D}}{W_{R\&D,m}}\right) ^{\frac{\epsilon }{1-\epsilon }}
\end{equation*}%
where $\Omega ={D^{\ast }}^{\omega }\cdot {D_{m}}^{\phi }$, $P_{J,m}$ is the
price of new designs and $W_{R\&D,m}$ is the CES wage index for the R\&D
sector:
\begin{equation*}
W_{R\&D,m}=\left( \sum_{r=1}^{Rm}\sum_{h=1}^{H_{r}} (b_{h,hi,r} \,
w_{h,hi,r})^{\frac{\sigma }{\sigma -1}}\right) ^{\frac{\sigma -1}{\sigma }}.
\end{equation*}

Note that given the constant return to scale technology of the R\&D sector,
the average cost corresponds to the marginal cost and there is no profit at
equilibrium, even in the short run. Moreover, as we assumed that new designs
were only used for one period, the R\&D sector does not benefit from rents
or royalties.

The demand of the R\&D sector for each variety of highly skilled labour from
region $q$ then takes the following form:

\begin{equation}
l_{R\&D}^{h,hi,r} = \left( \frac{w_{h,hi,r}}{b_{h,hi,r}^{\sigma} \,
W_{R\&D,m}}\right) ^{\frac{1}{\sigma -1}}L_{R\&D,m}  \label{L}
\end{equation}

\subsection{Durable goods firms}

Durable goods firms use the output of national R\&D firms and supply inputs
to final goods firms. In order to start
operating, the firm $v$ in the durable goods sector of region $r$ must
acquire one design and transform it into a new production process. The firm
can only obtain designs from its national R\&D sector by buying a licence
which must be renewed each period. Production also entails a fixed cost
denoted by $FC_{v,r}$. Finally, the firm receives subsidies from the
national government ($Sub_{m}^{v,r}$) and of the EU ($Sub_{EU}^{v,r}$). It
operates under monopolistic competition and produces one variety of durable
goods using physical capital. The production function is:
\begin{equation}
z_{v,r}=K_{v,r}  \label{Zvr}
\end{equation}

Capital is financed by selling assets $a_{v,r}$ to households on the $M$
national financial markets, which implies that $a_{v,r}=P_{r}^{k}K_{v,r}$,
with $P_{r}^{k}$ being the price of physical capital. Asset $a_{v,r}$ yields
a gross return $r_{v,r}^{k} \, P_{r}^{k}$ which corresponds to the rental
price for one unit of capital. We assume capital to depreciate at a rate $%
\delta_K$. This in fact corresponds to the mobile capital framework of \cite%
{MartinRogers95} which assumes that (i) capital is mobile between regions
and (ii) the revenue of capital is repatriated to the owner's region.

Each unit of capital is a CES aggregate of varieties of final goods bought
in all regions:
\begin{equation}
K_{v,r}=\left( \sum_{q=1}^{R}\sum_{s=1}^{S}
\sum_{i=1}^{N_{s,q}}(k_{v,r}^{i,s,q})^{\theta }\right) ^{\frac{1}{\theta }}
\end{equation}

This index is equivalent to the one representing preferences of households
which implies that price of capital is equal to the consumer price index,
i.e. $P_{r}^{k}=P_{r}^{c}$. Importantly, note that the price of capital is
region-specific. This reflects the fact that varieties constituting physical
capital must partly be imported. Given the existence of transport cost,
physical capital is more costly in small/peripheral regions.

Transforming designs into an effective new production process is uncertain.
We denote the probability to succeed in using a new design by $\phi $. In
order to capture the empirically well-documented fact that the capacity of a
region to innovate depends on its technological level and the skills
embodied in its human capital (see for instance \cite{RodriguezPoseCrescenzi08}),
we assume that $\phi $ depends on the existing stock of operational
processes which also corresponds to the number of durable goods firms, $A_{r}
$, and its stock of human capital, $HC_{r}$:
\begin{equation}
\phi _{r}=\left( \frac{A_{r}}{\sum_{r=1}^{R_{m}}A_{r}}\right) ^{\nu }\left(
\frac{HC _{r}}{\sum_{r=1}^{R_{m}}HC_{r}}\right) ^{1-\nu }.
\label{eq.probability.designs}
\end{equation}%
The regional stock of human capital is defined as the number of effective
units of high skilled labour available in region $r$, i.e $HC_{r} =
\sum_{h=1}^{H_{r}} b_{h,hi,r} \, l_{h, hi, r}$.

The expected profit of the durable goods firm then reads
\begin{equation}
\pi _{v,r}=\phi _{r} \, \left[ p_{v,r}^{z} \, z_{v,r}-r_{v,r}^{k} \,
P_{r}^{c} K_{v,r} - P_{J,m} - FC_{v,r} + Sub_{m}^{v,r} + Sub_{EU}^{v,r} %
\right]  \label{eq.profits.spec}
\end{equation}

Profit maximisation under the constraint (\ref{Zvr}) leads the durable goods
firm to address the following demand for each variety of final goods:
\begin{equation}
k_{v,r}^{i,s,q}=\left( \frac{\tau _{s,q,r} \left(1+t_{s,m}^{c}\right)
p_{i,s,q}}{\beta _{s}\cdot P_{r}^{c}}\right) ^{\frac{1}{\theta -1}}K_{v,r}
\label{K}
\end{equation}
The firm also sets its price as a mark-up over marginal cost with
\begin{equation}
p_{v,r}=\frac{MC_{v,r}}{\theta }  \label{PZ}
\end{equation}%
where $MC_{v,r}=r_{v,r}^{k}P_{r}^{c}$. This implies that production of the
durable goods firm and hence its demand for capital depends negatively on
the rental price of capital and positively on the demand addressed to the
firm (accelerator mechanism). Investment corresponds to the variation in the
stock of capital plus depreciation:
\begin{equation*}
I_{v,r}=\Delta K_{v,r} + \delta_K \, K_{v,r}
\end{equation*}%
It is financed by the issuance of new assets, i.e. $P_{r}^{c} \,
I_{v,r}=\Delta a_{v,r}$.

Note that, given the form of the production function (\ref{Zvr}) adopted for
the durable goods sector, the production function of a final goods firm
reads
\begin{equation}
y_{i,s,r}=A_{r}^{\alpha _{s}/\rho} \, K_{i,s,r}^{\alpha _{s}} \,
L_{i,s,r}^{1-{\alpha _{s}} }\,KG_{r}^{\alpha _{G}}-FC_{i,s,r}
\label{eq.value.added2}
\end{equation}
i.e. the volume of output depends of the use of capital (embodied in the
durable goods) and labour. As in \cite{Romer1990}, it also depends on
technological change which takes the form of an increase in the the range of
durable goods the firms have access to. This range is endogenous and
determined by the market conditions faced by durable goods firms, among
which the number of final goods firms on the regional market and the
performance of the national R\&D sector.

\section{Public sector}

\subsection{Government}

We assume a multi-level governance framework where the national government
interacts with the EU level. The expenditure of the national government of
country $m$ consists in consumption of final goods $GC_{m}$, transfers to
households $TR_{H,m}$, subsidies to firms $Sub_m$, and government investment
$GI_{m}$. These components of government expenditure are all assumed to be
fixed at exogenous levels, although they can serve as variables for
modelling policy shocks.

Let $G_{m}$ denote the sum of government consumption and investment. We
assume government consumption and investment to be distributed among the
regions of country $m$ according to the shares of the population. The
regional government also receives resources from the EU which we denote by $%
TR_{EU,q}$. The amount of public consumption and investment taking place in
region $q$ (assumed to be in country $m$) then reads\footnote{%
Note that by limiting public consumption and investment in a given region to
the allocation of resources received from the central government and the EU,
we rule out the possibility for regional governments to finance their
expenditure by raising their own taxes or issuing their own debt.}
\begin{equation*}
G_{q}= \frac{H_{q}}{ H_{m}} \cdot G_{m} + TR_{EU,q}
\end{equation*}

Analogously to households and firms, the regional governments have CES
preference defined over the set of varieties produced in the domestic
economy and abroad. We have
\begin{equation*}
G_{q}=\left(\sum_{r=1}^{R}\sum_{s=1}^{S}\beta _{s}\sum_{i=1}^{N_{s,r}}\left(
c_{G,q}^{i,s,r}\right) ^{\theta }\right) ^{\frac{1}{\theta }}
\end{equation*}

The demand addressed by the public sector of region $q$ to firm $i,s,r$ is
then
\begin{equation}
c_{G,q}^{i,s,r} = \left( \frac{\tau _{s,r,q} \left(1+t_{s,m}^{c}\right)
p_{i,s,r}}{\beta _{s} P_{q}^{c}}\right) ^{\frac{1 }{\theta-1}} G_{q}
\label{G}
\end{equation}

The government contributes to the EU budget and in particular to cohesion
policy funding, $CPF$, proportionally to its weight in the EU GDP:
\begin{equation*}
TR_{m,EU}=\frac{GDP_{m}}{GDP}\, CPF
\end{equation*}%
where $GDP_{m}=\sum_{r=1}^{Rm} GDP_{r}$ and $GDP=\sum_{m}GDP_{m}$. $GDP_{r}$
is GDP of region $r$ and is defined in the next section.

The government levies taxes on consumption as well as on capital and labour
income which constitutes its revenues:
\begin{eqnarray*}
T_{m} & = & \sum_{q=1}^{R_m}H_q \sum_{r=1}^{R} \sum_{s=1}^{S} t_{s,m}^{c} \,
N_{s, r} \, p_{i,s,r} \, \tau_{r,q,s} \, c_{h,q}^{i,s,r} \\
& + & t_{m}^{w} \left(\sum_{q=1}^{Rm} \sum_{e=lo,me,hi} \sum_{h=1}^{H_q }
w_{h,e,q} \, l_{h,e,q}\right) \\
& + & t_{m}^{\pi } \sum_{q=1}^{Rm} H_q \, KI_{h,q}
\end{eqnarray*}

The public deficit in country $m$ is the difference between government
expenditure, including interests on the outstanding debt, and revenue:
\begin{equation*}
D_{m}=\sum_{q=1}^{Rm} P_{q}^{c} \, G_{q} + TR_{H,m} + TR_{m,EU} + r_{m}^{G}
B_{G,m} + Sub_{m} - T_{m}-\sum_{q=1}^{R_m}TR_{EU,q}
\end{equation*}
where $B_{G,m}$ and $Sub_{m}$ are respectively the public debt and
government subsidies in country $m$, with\footnote{%
This formulation does not prejudge how subsidies change with the number of
firms. According to the policy scenario envisaged, the total amount of
subsidies could increase/remain constant with the number of firms while
subsidies allocated to individual firms remain constant/decrease with the
number of firms.}
\begin{equation*}
Sub_{m} =\sum_{r=1}^{Rm} \sum_{s=1}^{S} N_{s,r} \, Sub_{m}^{i,s,r} +
Sub_{m}^{R\&D} + \sum_{r=1}^{Rm} A_r Sub_{EU}^{v,r}
\end{equation*}

Finally, the stock of public capital in region $q$ increases with the level
of public investment of the regional government and decreases with
depreciation:
\begin{equation*}
\Delta KG_{q} = GI_{q} - \delta_K \, KG_{q}
\end{equation*}

\subsection{Modelling policy intervention}

In order to model the European cohesion policy (ECP) interventions, we
regroup the different ECP expenditure categories into 5 broader groups of
policy instruments (see Table \ref{tab.policy.interventions}). R\&D related
policy measures are modelled as subsidies ($Sub_{EU,q}^{R\&D}$) reducing
fixed costs in the R\&D sector. Policy instruments aimed at increasing human
capital are modelled as an education investment in skill-specific human
capital, $\Lambda_{e}$ and increases public consumption in the regions
benefiting from the intervention. Transport infrastructure investments are
modelled as a reduction of trade costs, $\tau _{s, r, q}$. Other
infrastructure investments are implemented in RHOMOLO as an increase of the
stock of public capital, $KG_{r} $. These interventions also increase the
level of public consumption. ECP policy measure affecting particular
industries or services are modelled as government subsidies reducing fixed
costs in the final goods and/or in the durable goods sector ($Sub_{EU,q}^{FG}
$ and $Sub_{EU,q}^{z})$. Finally, technical assistance is assumed to
increase government consumption.

\begin{table}[h]
\caption{Modelling of policy intervention in RHOMOLO}
\label{tab.policy.interventions}
\begin{center}
\begin{tabular}{lll}
\hline\hline
Field & Implementation in Rhomolo & Variables \\ \hline
RTD & Reduction of fixed costs in R\&D sector & $Sub_{EU,q}^{R\&D}$ \\
Human resources & Education investment in skill-specific human capital & $%
\Lambda _{e}$, $TR_{EU,q}$ \\
Infrastructure & Reduction of trade costs & $\tau _{s, q, r}$, $TR_{EU,q}$
\\
& Increase of the stock of public capital & $KG_{q}$, $TR_{EU,q}$ \\
Industry and services & Reduction of fixed costs in final goods sector & $%
FC_{i,s,q}$, $Sub_{EU,q}^{FG}$ \\
& Reduction of fixed costs in durable goods sector & $FC_{v,q}$, $%
Sub_{EU,q}^{z}$ \\
Technical assistance & Increase in public consumption & $TR_{EU,q}$ \\
\hline\hline
\end{tabular}%
\end{center}
\par
{\scriptsize Notes: The presented policy interventions are illustrative.
Many more policy instruments and their combinations can be implemented in
RHOMOLO.}
\end{table}

In order to translate the impact of a particular policy measure on the model
variables, we make use when relevant of complementary models or employ
estimates from the literature. For example, in order to simulate the TEN-T
investments in transport infrastructure, the improvements in the transport
network due to transport infrastructure investments are first simulated with
the transport model TRANSTOOLS, where the units of measurement are
kilometres of new infrastructure, number of additional lanes, maximum speed,
etc. In a second step, the impact of the changes in the accessibility of
regions on economic variables is simulated with RHOMOLO, where the units of
measurement are relative prices, wages, employment, GDP, etc.

In addition to supply-side effects, the ECP interventions have also
demand-side effects (see listed in Table \ref{tab.policy.interventions}).
Both the demand and supply side effects together with the induced general
equilibrium effects determine the net policy impact and hence all are
important for policy incidence. The demand-side effects are implemented as
additional government expenditure of final demand and investments goods.

\section{Market equilibrium and closure rules}

\subsection{Goods, labour and innovation markets}

All households and all firms within a given sector are assumed to be
symmetric, which implies that in a specific regions $r$ wages and quantities
consumed are identical for all households while prices and quantities
produced are identical for all firms.

The firm $i, s, r$ faces demand from four types of agents: households
(domestic and foreign) $D_{H}^{i,s,r}$, firms of the final goods sector $%
D_{F}^{i,s,r}$, firms of the durable goods sector $D_{K}^{i,s,r}$ and the
domestic public sector $D_{G}^{i,s,r}$:
\begin{eqnarray*}
D_{H}^{i,s,r} &=&\sum_{q=1}^{R} H_{q}\, c_{h,q}^{i,s,r} \\
D_{F}^{i,s,r} &=&\sum_{u=1}^{S}\sum_{q=1}^{R} N_{u,q} \, x_{j,u,q}^{i,s,r} \\
D_{K}^{i,s,r} &=&\sum_{q=1}^{R-1} A_{q}\, k_{v,q}^{i,s,r} \\
D_{G}^{i,s,r} &=&\sum_{q=1}^{R-1} c_{G,q}^{i,s,r}
\end{eqnarray*}

where $c_{h,q}^{i,s,r}$, $x_{j,u,q}^{i,s,r}$, $k_{v,q}^{i,s,r}$ and $%
c_{G,q}^{i,s,r}$ are respectively given by equations (\ref{eq.consumption2}%
), (\ref{F}), (\ref{K}) and (\ref{G}). The four components of total demand
feature the same price elasticity and the firm sets its price, $p_{i,s,r}$,
according to the rule given by equation (\ref{PF}), thereby equating demand
and supply:
\begin{equation*}
X_{i,s,r} = D_{H}^{i,s,r} + D_{F}^{i,s,r} + D_{K}^{i,s,r} + D_{G}^{i,s,r}
\end{equation*}

GDP of region $r$ then corresponds to $\sum_{s=1}^{S}N_{s,r}\cdot
P_{i,s,r}^{y}\cdot y_{i,s,r}=\sum_{s=1}^{S}N_{s,r}\cdot P_{i,s,r}^{y}\cdot
X_{i,s,r}$.

In region $q$, $H_q$ different varieties of low, medium and high skilled
labour are supplied on the labour market. Labour supply of skill level $e$
by on household $h$ in region $q$, denoted as $l_{h,e,q}$ is given by
equation (\ref{labour-supply}).

Labour demand stems from the final goods sector on the one hand and from the
national R\&D sector on the other hand. Labour demand from the final sector
is obtained by aggregating individual firms demand for for labour of skill
level $e$ and variety $h$, denoted by $l_{i, s, q}^{h,e}$, is given by
equation (\ref{l}). Labour demand from the national R\&D sector for highly
skilled labour of variety $h$ from region $q$ is denoted by $%
l_{R\&D}^{h,hi,q}$ and given by equation (\ref{L}).

Prices and quantities adjust so as to obtain equilibrium on the labour
market, i.e.:

\begin{eqnarray*}
l_{h,e,q} & = & \sum_{s=1}^{S} \sum_{i=1}^{N_{s,q}} l_{i, s, q}^{h,e} \qquad
\text{for} \, e = lo, me \\
l_{h,hi,q} & = & \sum_{s=1}^{S} \sum_{i=1}^{N_{s,q}} l_{i, s, q}^{h,hi} +
l_{R\&D}^{h,hi,q}
\end{eqnarray*}

On the market for durable goods of region $r$, the firm $v$ faces the
following demand:
\begin{equation*}
D_{F}^{v,r} = \sum_{s=1}^{S} N_{s,r} \cdot z_{i,s,r}^{v,r}
\end{equation*}
where $z_{i,s,r}^{v,r}$ is specified by equation (\ref{Z}). The price
setting rule (\ref{PZ}) ensures that supply equals demand so that
\begin{equation*}
z_{v,r} = D_{F}^{v,r}
\end{equation*}

Finally, the demand for designs addressed to the R\&D sector corresponds to
the number of firms willing to operate in the durable goods sector $%
\sum_{r=1}^{Rm}N_{r}^{z}$. As described in the next section, this number
depends on the price of designs, $P_{J,m}$, so that at equilibrium we have
\begin{equation*}
J_{m}^{r} = \sum_{r=1}^{Rm} \frac{A_{r}}{\phi _{r}}
\end{equation*}

\subsection{Financial markets}

We select a saving driven closure rule where private saving is determined as
a constant fraction of households' income (see equation \ref{BC}). At
equilibrium, (i) private saving must finance private investment, public
deficits and the deficit of the trade balance; and (ii) returns on the three
types of assets held by households must be equal. Finally, we assume that
financial markets are fully integrated at the level of the $m$ countries.

Private investment in region $r$ is the sum of investment of firms of the
durable goods sector (i.e. the firms directly using capital as a production
factor): $P_r^C \, I_{r} = \sum_{v=1}^{A_r}{P_r^C \, I_{v,r}} = A_r \, P_r^C
\, I_{v,r}$.

The trade balance deficit of each country ($TB_{m}$) corresponds to the
value of its exports minus the value of its imports, $TB_{m}=X_{m}-M_{m}$
where:
\begin{eqnarray}
X_{m} &=&\sum_{r=1}^{Rm}\sum_{s=1}^{S-1}\sum_{i=1}^{N_{s,r}} \tau _{S, r, R}
\, p_{i,s,r} \, c_{R}^{i,s,r}  \label{eq.exports.equilibrium} \\
M_{m} &=&\sum_{r=1}^{Rm}\sum_{h=1}^{H_{r}} \tau _{S,R,r} \, p_{S,R} \,
c_{h,r}^{S,R}  \label{eq.imports.equilibrium}
\end{eqnarray}

The trade balance of the domestic economy then corresponds to the sum of the
national trade balances with respect to the rest of the world:
\begin{equation*}
TB=\sum_{m=1}^{M} TB_{m}
\end{equation*}

We therefore have
\begin{equation*}
S = \sum_{r=1}^{R-1}\sum_{h=1}^{H_r}{S_{h,r}} = \sum_{r=1}^{R-1}{A_{r} \,
P_r^C \, I_{r}} + \sum_{m=1}^{M}{D_m} + TB
\end{equation*}

Finally, arbitrage on the financial markets equalises net returns on
financial assets. The net return for holding capital in firm $v,r$ is $%
(r_{v,r}^k - \delta_K) P_r^C + (1-\delta_K) \Delta P_r^C$. Firms are
symmetric and hence $r_{v,r}^k = r_{r}^k$ for all $v$. Letting $r_{G,m}$
denote the return on government bonds of country $m$ and $r_F$ the return on
foreign bonds, the arbitrage condition is
\begin{equation*}
(r_{r}^k - \delta_K) P_r^C + (1-\delta_K) \Delta P_r^C = r_{G,m} = r_F
\end{equation*}
for all $m = 1 \ldots M$ and for all $r = 1 \ldots R-1$. Note that the
required gross return for physical capital $r_{r}^k \, P_r^C$ is higher in
regions where the price of capital $P_r^C$ is high. This reflects the fact
that depreciation incurs a higher financial loss when the resources needed
to acquire capital are more important, which is for instance the case in
remote regions.

Households hold assets in proportion to their saving. The accumulation of
assets by household $h, q$ is then described by the following law of
motions:
\begin{eqnarray*}
\Delta B_{h,q}^{k,v,r} & = & \frac{S_{h,q}}{S} \, \Delta a_{v,r} \\
\Delta B_{h,q}^{G,m} & = & \frac{S_{h,q}}{S} \, D_{m} \\
\Delta B_{h,q}^F & = & \frac{S_{h,q}}{S} \, TB
\end{eqnarray*}
where $S = \sum_{q=1}^{R-1} \sum_{h=1}^{H_q} S_{h,q}$ corresponds to the
total savings of domestic households.

\section{Location and spatial equilibrium}

\subsection{Why does space matter in RHOMOLO?}

The model breaches a number of the conditions identified by \cite{Starrett78}
for having perfectly homogenous distribution of economic activity in space.
In particular, agents and factors of production are partly immobile,
locations are not uniforms (because population and accessibility varies from
one regions to the other), the economy is open, there is imperfect
competition on product and labour markets and the introduction of knowledge
spill-over makes some markets incomplete. There are however two sets of
elements without which the issue of location and space would not exist in
the model: the combination of trade cost and increasing returns on the one
hand, and the combination of knowledge spill-over and localised
externalities on the other hand.\footnote{%
See \cite{agglomeration2013} for a detailed description of agglomeration and
dispersion forces and mechanisms in RHOMOLO.}

Both consumers and producers face positive \textit{trade costs} for
importing final/investment goods and intermediate inputs. On the consumer
side, trade costs enter the consumer price index (\ref%
{eq.consumer.price.index}). On the producer side, trade costs enter the
intermediate goods price index (\ref{eq.intermed.price.index}) and the
investment price index (similar to the consumer price index). However,
departing from the standard framework of the new economic geography
literature, bilateral trade costs between regions are assumed to be
asymmetric and the internal trade costs to be positive. Values for the
inter-regional trade costs come from the data, instead of being calibrated
or proxied by distance \citep[see][for
details]{IvanovaKancsStelder11,rhomolodatapaper,rhomolosams}.

\textit{Increasing returns to scale} are introduced via fixed costs in firms
production functions (\ref{eq.value.added}) and (\ref{eq.profits.spec}).
Following \cite{Venables96}, they are made of part of the firms output. In
contrast to trade costs, fixed costs is strictly speaking an exogenous
variable rather than a parameter. Nevertheless, they can be used to
calibrate the model.

The combination of increasing returns, - preventing the endless division of
the scale of economic activities and hence the emergence of so-called
backyard capitalism -, and of transport cost, - without which the issue of
space would be irrelevant -, makes access to large markets a determinant of
the firms performance. Access to large markets allows the exploitation of
economies of scale and hence increase profits. Location (close to a large
market) then becomes a decision variable.

\textit{Localised externalities} enter RHOMOLO trough technological and
knowledge spillovers whose scope is assumed to be limited in space to the
boundary of the region. Indeed, localised externalities are region-specific
and determine the relationship between the density of workers and durable
goods firms in a region on the one hand and the performance of the local
durable goods sector and hence the productivity of factors on the other
hand. When the number of durable goods firms increases in one region,the
total productivity of factors used in the industry also increases. This
leads to an increase in the number of firms in the final goods sector which
in turn increases demand for durable goods and hence profits in the durable
goods sector. This type of Marshallian externality (see for instance \cite%
{Marshall1890} or \cite{Scitovsky54}) implies that R\&D and technological
progress tends to be spatially concentrated in a limited number of places.

\subsection{Spatial equilibrium}

In the short run, pure profit may exist. However, in the long run, this will
trigger the entry of new firms on the market which will decrease the demand
addressed to each firm and hence reduce the level of profit.\footnote{%
The expressions describing total demand are relatively complicated but one
can indeed show that it is a decreasing function of the number of firms. In
the simple case where there is only one sector and one region, the demand
addressed to a particular firm by consumers is $1/N\cdot I/p$ where $I$ is
the income devoted to consumption} This process takes place until pure
profit is completely exhausted. The profit of firm $i,s,r$ reads
\begin{eqnarray}
\pi _{i,s,r} &=&p_{i,s,r} \, X_{i,s,r}-P_{i,s,r}^{y} \,
y_{i,s,r}-\sum_{u=1}^{S}P_{r}^{u} \, X_{i,s,r}^{u}-P_{i,s,r}^{y} \, FC_{r}
\notag \\
&=&p_{i,s,r} \, X_{i,s,r}-P_{i,s,r}^{y} \, X_{i,s,r}-\sum_{u=1}^{S}{\
a_{s}^{u-1}} \, P_{r}^{u} \, X_{i,s,r}-P_{i,s,r}^{y} \, FC_{r}
\label{eq.profit1}
\end{eqnarray}

Pure profit is equal to zero when the price equals average cost, i.e.
\begin{equation}
0=p_{i,s,r} - P_{i,s,r}^{y}-\sum_{u=1}^{S}{{a_{s}^{u-1} } \, P_{r}^{u}}%
-P_{i,s,r}^{y} \, FC_{r}/X_{i,s,r}  \label{eq.profit2}
\end{equation}

Using the price setting rule (\ref{PF}), one obtains the level of production
corresponding to zero profit:
\begin{equation*}
X_{i,s,r}^{\ast }=\frac{P_{i,s,r}^{y} \, FC_{r}}{\frac{1-\theta }{\theta }
\, \left[ P_{i,s,r}^{y}-\sum_{u=1}^{S}{a_{s}^{u}}^{-1} \, P_{r}^{u} \right] }
\end{equation*}

The same mechanism applies to the durable goods sector. For each firm of the
sector, pure profit is exhausted when demand is such that the price it sets
is equal to average cost:
\begin{equation*}
p_{v,r} = r_{v,r}^k \, P_r^C + P_{J,m}/ z_{v,r} + FC_{v,r}/ z_{v,r}
\end{equation*}

By equation (\ref{PZ}), the price is a mark-up over marginal cost which,
combined to the expression above, gives the production level which
annihilates pure profit:
\begin{equation*}
z_{v,r}^{*} = \frac{ P_{J,m} + FC_{v,r}}{\frac{1-\rho}{\rho} \, \left[%
r_{v,r}^k \, P_r^C \right]}
\end{equation*}

We then have a system of $s \times r$ equations of the type $X_{i,s,r}^{*} =
D_{H}^{i,s,r} + D_{F}^{i,s,r} + D_{K}^{i,s,r} + D_{G}^{i,s,r} $ plus $r$
equations $z_{v,r}^{*} = D_{F}^{v,r}$ with $s \times r + r$ unknowns
corresponding to the long term number of firms in each sector and in each
region, $N_{s,r}^*$ and $A_{r}^*$.

Transition to the long term number of firms is not immediate and is
described by the following law of motion, which is assumed to be the same in
every region and sector: $\Delta N=\lambda \cdot (N-N^{\ast })$. The number
of firms in each region determines the spatial distribution of economic
activity in model. It is fully endogenous and incorporates several
agglomeration and dispersion forces.

\subsection{Agglomeration and dispersion forces}

Four effects drive the mechanics of endogenous agglomeration and dispersion
of economic agents in RHOMOLO: the \textit{market access effect}, the
\textit{price index effect}, the \textit{market crowding effect} and the the
\textit{localised externalities effect}.

The \textit{market access effect} is based on the fact that, due to the
presence of increasing returns and transport costs, firms in large/central
regions tend to have higher profits than firms in small/peripheral regions.
Firms therefore prefer to locate in large/central regions and export to
small/peripheral regions. Due to positive trade costs, the demand for a
region's output increases with its relative accessibility and its economic
size. This can be seen in equations (\ref{eq.consumption2}), (\ref{F}), (\ref%
{K}) and (\ref{G}), according to which total demand addressed to firm $i, s,
r$, and hence its profit decreases with trade costs, $\tau _{s, r, q}$,
decrease with an elasticity $\frac{1}{1-\theta}$. The weighted average trade
costs can be lower either due to large internal market (low value of $\tau
_{s, r, r}$) or due to good accessibility of/central location of a region
(low value of $<\tau _{s, r, q}$), or both.

The profitability of firms facing larger demand is enhanced due to the
existence of increasing returns, as growth in output reduces the average
production costs. This can be seen by combining equations (\ref%
{eq.prod.function}), (\ref{eq.value.added}) and (\ref{eq.profit1}),
according to which an increase in output, $X_{i, s, r}$, reduces the share
of fixed costs in average costs, and hence increases the firm's profit.

The \textit{price index effect} describes the impact of firms' location and
trade costs on the cost of intermediate inputs and of durable goods for
producers of final demand goods. This follows the vertical linkage framework
of \cite{Venables96}. Large/central regions with more firms import a
narrower range of products, which reduces trade costs. Therefore,
intermediate inputs are less expensive in large/central regions than in
small/peripheral regions. This can be seen in the intermediate inputs price
index (\ref{eq.intermed.price.index}) which decreases in trade costs with
elasticity $1$. This suggests that that total trade costs, $%
\sum_{r=1}^{R}\tau _{s, r, q}$, and hence the cost of production is lower in
large/central regions. Moreover, production costs are also lower in regions
with a large number of durable goods firms. Indeed, one easily checks that
the durable goods price index (\ref{eq.specialis.price.index}) decreases
with the number of durable goods firms $A_{r}$. Because of lower production
cost, firms purchasing intermediate inputs and using durable goods as a
factor of production would prefer to locate in large/central regions.

The \textit{market crowding effect} capture the fact that, because of higher
competition for input and output markets, firms prefer to locate in
small/peripheral regions with fewer competitors than in large/central
regions where competition is fiercer. Indeed, when the number of firms in
large/central regions increases consumption of differentiated goods is
fragmented over a larger number of varieties, implying that each firm's
output and profit decreases. Given that the entry of new firms has a
negative effect on profitability of incumbents in large/central regions,
this \textit{market crowding effect} works as a dispersion force.

The effect of competition on output markets can be seen in equations (\ref%
{eq.consumption2}) and (\ref{eq.consumer.price.index}), according to which
the demand of output produced by firm $i, s, r$ is decreasing in the number
of final goods firms. Lower output, and hence profit, gives the incentive to
firms to move away from large/central regions to small/peripheral regions
with fewer competitors.

The effect of competition on input markets works through prices of spatially
immobile production factors, namely labour and durable goods. Agglomeration
of firms in large/central regions bids up prices for such production factors
which reduces the incentive to locate in places where the number of firms is
large.

The \textit{local externalities effect} works through the probability to
succeed in transforming designs into a new production process, $\phi _{r}$,
which depends on the pre-existing regional stock of durable goods firms, $%
A_{r}$, and the stock of human capital, $HC_{r}$ (see equation \ref%
{eq.probability.designs}). In particular, the probability to operationalise
a new design is higher in regions where the number of durable goods firms is
large. As a result, the accumulation of technology is facilitated in regions
largely endowed with durable goods firms, creating the conditions for R\&D
and technological progress to agglomerate in places where the stock of
knowledge and of technology is already large.

The table below summarises the endogenous location mechanisms which drive
the geographical distribution of final goods and durable goods firms and
foster their agglomeration or dispersion in space.
\begin{table}[h]
\caption{Agglomeration and dispersion in RHOMOLO}
\label{tab.agglomeration.mechanisms}
\begin{center}
\begin{tabular}{lccc}
& Final goods firms & durable goods firms &  \\ \hline
\textit{Market access effect} & $\Uparrow$ & $\Uparrow $ &  \\
\textit{Price index effect} & $\Uparrow$ &  &  \\
\textit{Market crowding effect} & $\Downarrow$ & $\Downarrow $ &  \\
\textit{Local externalities effect} & $\Uparrow$ & $\Uparrow$ &  \\ \hline
\end{tabular}%
\end{center}
\par
{\scriptsize Note: $\Uparrow $ denotes agglomeration, $\Downarrow $ denotes
dispersion.}
\end{table}

Note that the agglomerations of final goods and of durable goods firms
reinforce each other. A large number of final goods firms means a large
market for durable goods firms which enhances the market access effect for
the durable goods sector. A large number of durable goods firms implies that
a large number of varieties of durable goods are available for final goods
firms which enhances the price index effect for the final goods sector.

\section{Conclusion}

Cohesion policy shifts the spatial equilibrium at the regional level within
the EU and the Member States by increasing the capacity for growth in the
regions that are lagging behind and to some extent also by mobilising the
unused capacity in other regions. It does so by supporting investments in
the trans-European infrastructure networks connecting the regions as well as
by stimulating measures fostering the development of human resources,
research and innovation and, in general, improving the standard of living
and attractiveness of the regions. Although the room for public funding and
redistribution is limited by balanced budget requirements, the impact on the
less developed regions can be very substantial if the forces of
agglomeration and dispersion of economic activity, as they are laid out in
the New Economic Geography literature, are taken into account.

This paper presents a spatial general equilibrium framework in which the
interplay of agglomeration and dispersion forces, including the ones set in
motion by cohesion policy can be analysed in a novel and theoretically
consistent way, including the impact in the net contributing Member States.
Particular attention is paid to income and capital movements within and
between regions that are generated by the stimulus to the regions. This will
allow an assessment of the feedback to the Member States and regions and the
possibility that in the longer run they will all benefit from the additional
growth that is generated.

The paper carefully analyses the implications of cohesion policy
interventions on the spatial equilibrium in terms of income and employment.
In doing so, it sheds new light on how the success of cohesion policy can be
measured. The paper recognises the limitations of a comparative static
approach and advocates further work and extensions of the model and its
potential use in the direction of dynamics, in particular by incorporating
the results of research on long-term productivity developments and migration
between regions.

\singlespacing
\setlength{\bibsep}{5pt} 

\section*{References}

\bibliographystyle{ecca}
\bibliography{rhomolo2}
\onehalfspacing

\end{document}